\documentclass[aps,prb,twocolumn, groupedaddress, showpacs]{revtex4}
\usepackage{amsmath}
\usepackage{amssymb}
\usepackage{graphicx}
\usepackage{color}
\definecolor{Blue}{rgb}{0.3,0.3,1}
\usepackage{tikz}
\usepackage{pgffor}
\usepackage{verbatim}
\begin{document}

\title{Rectification in Y-junctions of Luttinger liquid wires}

\author{Chenjie~Wang and D.~E.~Feldman}
\affiliation{Department of Physics, Brown University, Providence, Rhode Island 02912, USA}

\date{\today}

\begin{abstract}
We investigate rectification of a low-frequency ac bias in Y-junctions of one-channel
Luttinger liquid wires with repulsive electron interaction. Rectification emerges due to three
scatterers in the wires. We find that it is possible to achieve a higher rectification current in a Y-junction
than in a single wire with an asymmetric scatterer
at the same interaction strength and voltage bias. The rectification effect is the strongest in the absence of the time-reversal symmetry.
In that case, the maximal rectification current
can be comparable with the total current $\sim e^2V/h$ even for
low voltages, weak scatterers and modest interaction strength.
In a certain range of low voltages, the rectification current can grow as the voltage decreases. This leads to  a bump in the $I$-$V$ curve.

\end{abstract}



\maketitle

\section{Introduction}

Recently there was much interest in rectification in nanoscale systems.
\cite{1a,2a,3a,CB,4a,5a,6a,7a,8a,asym-scat,9a,y-carbon,10a,11a,12a,SB,SZ,FSV,BFM,BFF,scheid1,scheid2}
Nonlinear mesoscopic transport exhibits much interesting physics such as a peculiar magnetic field dependence of the current \cite{SB,SZ} and negative differential resistance for the rectification current at low voltages.\cite{FSV,BFM}
Another motivation for the
investigation of mesoscopic diode or ratchet \cite{HM} effect comes from possible practical applications in nanoelectronics and energy conversion.
Following the pioneering paper by Christen and B\"uttiker \cite{CB} most attention has focused on a simpler case of a Fermi-liquid conductor. At the same time, the diode
effect requires a combination of spacial asymmetry and strong electron interactions in the conductor. Hence, one may expect a stronger ratchet current in strongly interacting Luttinger liquid systems. This expectation has been confirmed by a recent study of transport asymmetries in one-channel quantum wires.\cite{FSV,BFM,BFF} Refs.~\onlinecite{FSV,BFM,BFF} have focused on a one-channel Luttinger liquid in a linear conductor in the presence of a single asymmetric scatterer. This is the conceptually simplest situation
giving rise to rectification.  At the same time, changing geometry may increase asymmetry and hence the rectification current. In this paper we consider an
asymmetric setup based on a Y-junction of three quantum wires with three impurities. We show that a stronger diode effect can be achieved in such system than in a linear
Luttinger liquid and rectification is possible even in the case of symmetric point scatterers.

We focus on the simplest one-channel Y-junctions. More complicated Luttinger liquid junctions, such as Y-junctions of single-wall carbon nanotubes, are also of interest. In particular, it might be easier to make such junctions in a reproducible way.

Y-junctions are among basic elements of electric circuits, however, a theoretical investigation of Luttinger liquid Y-junctions\cite{add1,1,2,3,4,5,6,7,8,add2,9,10,11,12,13,14,15,16,17,18}
has begun only during the last decade. By now, there is a good understanding of linear conductance near various fixed points as well as tunneling density of
states.\cite{add2,13}
In this paper we extend the previous research to the problem of transport asymmetries.
Specifically, we consider a setup of the type shown in Fig. 1. We assume that one of the three terminals is kept at zero voltage. ac voltages with amplitudes $V$ and $\gamma V$, $\gamma\sim 1$, are applied to the remaining two terminals. In a general case, a dc-current is generated in each of the three wires in the junction, Fig. 1. The three currents are different but can be computed in a similar way. We calculate such rectification dc currents for various types of Y-junctions.
We focus on the limit of a low-frequency ac voltage bias. In order to determine the amplitude of the rectification effect in that limit, it is sufficient to
find the difference of the dc currents at the opposite dc bias voltages, i.e., compare the current when the potentials at the terminals are time-independent and equal 0, $V$ and $\gamma V$ with the current when the potentials are 0, $-V$ and $-\gamma V$ (cf. Refs.~\onlinecite{FSV,BFM}). This corresponds to a dc current generated by low-frequency square voltage waves.
We find that in some classes of Y-junctions the rectification current is higher than in a linear wire with the same strength of the repulsive electron interaction
at the same voltage bias.
In a certain interval of low voltages the rectification current exhibits a power-law dependence on the bias: $I\sim V^z$. The exponent $z$ can be negative. Clearly, such dependence
with a negative exponent cannot extend all the way to zero voltage as $I=0$ at $V=0$. Hence, the rectification current reaches a maximum at a certain voltage.
We demonstrate that in junctions without time-reversal symmetry, the maximal rectification current can be comparable with the total current $\sim e^2V/h$
even for low voltages and modest interaction strength in the wires. In particular, such rectification current can be achieved in the ``island setup'',
illustrated in Fig.~1, which can be experimentally realized in quantum Hall
systems \cite{10}. For comparison, in a linear wire, the rectification current \cite{FSV,BFM}
is always much lower than $e^2V/h$.
The diode effect in Y-junctions is not as strong in the presence of time-reversal symmetry
as in its absence. Still, the maximal rectification current is higher than in a linear conductor.

The paper is organized as follows: first, after a brief qualitative discussion, we describe our setup and formulate a model. We then review the properties of Y-junctions and derive a general expression
for the rectification current. Next, we apply the general formalism to the model of three weakly connected wires. Then we determine the leading contributions to the rectification current at different interaction strengths and matching conditions at the junction. We discover several regimes with different voltage dependences of the current. Finally, we discuss how to build a junction with the maximal rectification current.

\section{Rectification in mesoscopic conductors}

\begin{figure}
\centering
\includegraphics[width=2in]{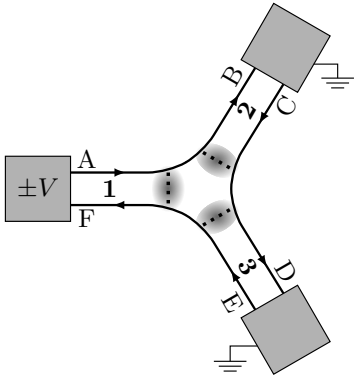}
\caption{Schematic picture of a Y-junction of quantum wires with three impurities. Voltage bias $\pm V$ is applied to the first wire. Wires 2 and 3
are connected to the ground. We calculate the dc current in wire 1. One can view AB, CD and EF as edges of an integer quantum Hall liquid. This corresponds to an
``island junction''.\cite{10}}\label{fig1}
\end{figure}

We consider a junction of $N$ one-channel wires.  In the subsequent sections we will specialize to the case of $N=3$ and assume that the system is spin-polarized or, equivalently,
that charge carriers are spinless fermions.
In all cases we assume that long range Coulomb interaction in the conductor is screened by the gates. In such situation the applied voltage bias
affects not only the current but also the charge density in the system.

Rectification is possible, if
a change of the voltage bias sign results in a change of not only the sign but also absolute value of the current. The latter obviously requires left-right asymmetry in the system.
Asymmetry may have two origins: asymmetry caused by the way how the bias is applied and geometric asymmetry due to, e.g., an asymmetric scatterer.\cite{FSV,BFM} In particular,
one can consider a situation in which several terminals are kept at zero electric potential and the potential of the other terminals changes between $+V$ and $-V$.
The charge density is different for the opposite voltage signs. This in turn affects the current and leads to rectification. Such density-driven rectification
is possible in a broad range of situations including systems without electron interaction in the presence of time-reversal symmetry. It has been investigated
in the context of a carbon nanotube junction in the
noninteracting-electrons approximation in Ref.~\onlinecite{y-carbon}. At a high voltage bias a strong rectification effect was found. In this
paper, on the other hand, we focus on the low-voltage regime and, in particular, the
universal behavior at low bias. In that limit the density-driven rectification effect is of little importance as
it results in a small rectification current $\sim V^2$ in a noninteracting system. Besides, the density-driven rectification requires a specific way to apply bias.
If we are interested, for example, in the transformation of incoming electromagnetic radiation into a dc current then clearly the potential oscillates in all terminals.
Thus, in this paper we focus on the rectification mechanism due to geometric asymmetry.

Y-junctions have a ``built-in'' geometric asymmetry. Indeed, let us focus on the current $I_1$ in one of the three wires connected by the junction. The current
$I_1=I_2+I_3$ equals the sum of the currents in the other two wires. Thus, the same current
enters the junction through two wires and leaves through one only. This means that the two sides of the junction are not equivalent since they correspond to 1 and 2 wires. Thus,
we may expect that geometric mechanism of rectification applies to any Y-junction.
However, as we demonstrate below, the geometric mechanism only works, if electron interaction is present or time-reversal symmetry is broken.

Indeed, let us consider a time-reversal invariant system of non-interacting spinless fermions in an $N$-terminal mesoscopic junction.
We calculate the current between terminals $1,\dots, K$ and $K+1,\dots, N$. We compare the current in the situation
when terminals $1,\dots, K$ are kept at the voltage $V$ and terminals $K+1,\dots,N$ are kept at zero bias (case 1) with the current in the situation when terminals $1,\dots,K$
are kept at zero bias and terminals $K+1,\dots, N$ are kept at the voltage $V$ (case 2).
For non-interacting fermions the current reduces to the sum of the single particle contributions corresponding to each energy in the window $0<E<V$. We thus compare such contributions for two opposite voltage biases. The wave function of an electron incoming from terminal $l$ with the momentum $k$ is $\psi_{k,l}=\exp(-ikx_l)+\sum_{m=1}^NS_{lm}\exp(ikx_m)$, where $x_m>0$ is the coordinate in wire number $m$ and $S_{lm}$ is a unitary scattering matrix. The above notations for the wave function imply that the probability to find
an electron with the wave function $\psi_{k,l}$ in wire number $m\ne l$ is determined by the outgoing wave and is proportional
to $|S_{lm}|^2$. For $l=m$, the probability is determined by both the incoming and outgoing waves.
The time-reversal symmetry implies that $\psi^*_{k,l}=\exp(ikx_l)+\sum_{m=1}^NS^*_{lm}\exp(-ikx_m)$ is also a solution of the Schr\"odinger equation.
The solution is made of $N$ incoming waves and one outgoing wave.
It can be represented as a linear combination of waves $\psi_{k,m}$ with
different $m$.
Hence, the outgoing wave satisfies the equation $\exp(ikx_l)=\exp(ikx_l)\sum_{m=1}^NS^*_{lm}S_{ml}$.
The absence of outgoing waves in the channels with numbers $n\ne l$ implies that
$\sum_m S_{lm}^*S_{mn}=0$.
Hence, $S^*=S^{-1}=S^\dagger$, i.e., $S$ is a symmetric matrix. Now we can compare the contributions to the current from particles with the energy $E=\hbar^2k^2/2m$
for two opposite signs of the bias. Electrons, incoming from different terminals, are not coherent. Hence, in case 1, $I_1\sim \sum_{l=1}^K\sum_{m=K+1}^N|S_{lm}|^2$.
In case 2, $I_2\sim \sum_{l=1}^K\sum_{m=K+1}^N|S_{ml}|^2$. The currents are equal from the symmetry of the scattering matrix and hence the diode effect is absent.

\begin{figure}
\centering
\includegraphics[width=3.4in]{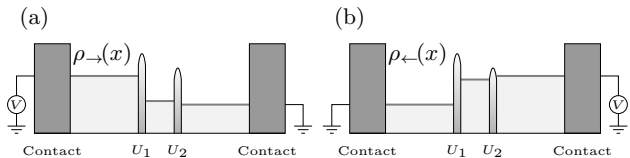}
\caption{A schematic picture of a linear wire with two scatterers of unequal strength. Charge density, averaged over Fiedel oscillations, follows a ``staircase'' profile. The direction of the staircase depends on the voltage sign as seen from figures 2a and 2b.}\label{fig2}
\end{figure}

In the presence of electron interaction, the rectification effect becomes possible. This can be understood already from a simple model with two point scatterers of unequal strength in a linear wire (Fig. 2). Similar rectification mechanisms operate in more complex junctions of Luttinger liquids.
We will assume that long-range Coulomb interactions are screened by the gates and thus the charge density depends on the voltage. In Fig. 2a we consider the situation with the incoming current from the left. Backscattering off two impurities results in a ``staircase'' charge density profile (we have averaged over Friedel oscillations). The ``staircase'' goes up as one moves from the right to the left. For the opposite voltage sign, the incoming current arrives from the right, Fig. 2b. We again have a ``staircase'' charge density profile but now the ``staircase'' goes down as one moves from the right to the left. In the absence of electron interactions, the transmission coefficients must be the same in both cases and rectification is absent. Let us now consider the effect of electron interactions in a simple mean-field Hartree picture. Since long-range Coulomb interactions are screened, the relation between the charge density $\rho(x)$ and the electric potential $W(x)$ is local.
Assuming small charge density variations, we thus find $W(x)\sim\rho(x)$.
Incoming electrons are scattered by a combined potential of the two impurities and  the Hartree potential $W(x)$. The latter depends on the incoming charge density and hence the applied voltage. Hence, for opposite voltage signs, electrons feel different backscattering potentials. This results in different transmission coefficients and thus rectification.

\begin{figure}
\centering
\includegraphics[width=3.4in]{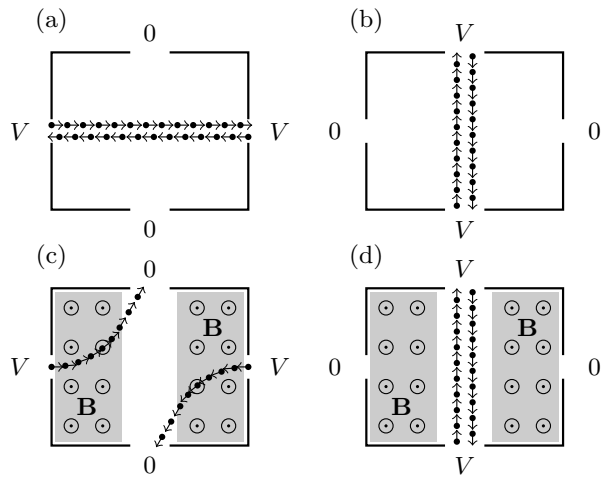}
\caption{Charge transport through the box exhibits no asymmetries at zero magnetic field, Figs. 3a,b, and is asymmetric at a finite field in shaded areas, Figs. 3c,d.}\label{fig3}
\end{figure}

Another mechanism of transport asymmetries comes from time-reversal symmetry breaking. The latter is possible in the presence of a magnetic field. We illustrate asymmetric transport in the absence of the time-reversal symmetry with a model system depicted in Fig. 3. While it is not a realistic model of any mesoscopic conductor, it is very simple and exhibits strong transport asymmetries. We assume no electron interactions in the model.

We consider a box with four holes in its sides. Each hole corresponds to a terminal in a more realistic description of a junction. We assume that all charge carriers, entering the box through the holes, have exactly the same speed, perpendicular to the wall in which the hole is made. The speed $v=\sqrt{-2meV}$, where $V$ plays a role, similar to the electrostatic potential of a terminal in a realistic junction. At $V=0$ the speed $v=0$ and hence no current is injected. In a Y-junction, charge density, injected from each terminal, depends on its electrostatic potential. A comparison between opposite voltage signs in a realistic junction corresponds in our model to the comparison between the situations with the charge injected through the right and left holes (Figs. 3a,c) and through the upper and lower holes
(Figs. 3b,d). In the absence
of a magnetic field, the current through each hole is zero in both situations: currents of the particles, injected through the opposite holes, cancel as shown in Figs. 3a,b. Let us now turn on a magnetic field in shaded areas (Figs. 3c,d).
If the charge carriers are injected from the upper and lower holes then the current remains zero even in the presence of a magnetic field (Fig. 3d). Indeed, the carriers never enter the region with the field. However, if the carriers arrive from the left and right holes then they are deflected by the field (Fig. 3c) and the electric currents are nonzero in all four holes. This illustrates
how magnetic field can result in transport asymmetry.

The above arguments make it plausible that the rectification effect is the strongest if both  strong electron interaction is present and time-reversal symmetry is broken.
This is confirmed by our calculations below for Y-junctions of quantum wires with and without time-reversal symmetry.

\section{Y-junctions}

In this section we formulate our model and review basic properties of Y-junctions.

We consider a Y-junction with the action

\begin{equation}
\label{L}
L=\int dt [\sum_{k=1,2,3}L_k-\sum_{k=1,2,3} T_k],
\end{equation}
where $L_k$ are the actions of three uniform wires and $T_k$ describe three impurities in the wires close to the junction point $x=0$. The above action must be supplied by matching conditions for three wires. They will be discussed below.
The action of a uniform wire is given by the equation

\begin{align}
\label{Lk}
L_k=  \int_{0}^{\infty} dx & [i\psi_I^{k\dagger}(\partial_t-v_F\partial_x)\psi^k_I+
i\psi_O^{k\dagger}(\partial_t+v_F\partial_x)\psi^k_O\nonumber\\ & -v_F\lambda(\psi_I^{k\dagger}\psi^k_I+\psi_O^{k\dagger}\psi^k_O)^2],
\end{align}
where $\psi^k_{I,O}$ are the operators of incoming and outgoing chiral electron fields, $v_F$ is the effective Fermi velocity, $v_F\lambda$
defines the interaction strength. We set $\hbar=1$ and the electron charge $e=1$ in most of the following text. We assume that the long range part of the Coulomb force is screened by the gates. Hence, the electron
density depends
on the voltage bias. The impurity Hamiltonians
\begin{equation}
\label{Tk}
T_k=\int_0^\infty dx U_k(x) [\psi^{k\dagger}_I\psi^k_O+h.c.],
\end{equation}
where $U_k(x)$ is the potential of the impurity in wire number $k$,
the impurity being located
close to the junction at $x=0$. We only take the backscattering part of the impurity Hamiltonian
into account as the forward scattering terms do not affect our results. We assume the same interaction strength and Fermi velocity in each wire.

In order to treat the case of strong interaction it is convenient to bosonize \cite{boson} the action in terms of the chiral fields $\phi_{O, I}^k$ such that
$\psi^k_{O/I}=F^k_{O/I}\exp(\pm ik_Fx+i\phi^k_{O/I})$, where the opposite signs should be chosen for the in- and out-fields, the commutator
\begin{equation}
\label{comm}
[\phi^k_{O/I}(y),\partial_x\phi^l_{O/I}(x)]=\mp2\pi i\delta(y-x)\delta_{kl}
\end{equation}
 and
$k_F$ plays the role of the effective Fermi momentum and determines the average charge density in the wires.
$F_{I,O}$ are Klein factors, necessary to ensure the proper Fermi commutation relations.
The local densities of in- and out-moving particles in point $x$ are $\rho_{O/I}=(k_F\pm\partial_x\phi_{O/I})/2\pi$.
The action now assumes the form
\begin{align}
\label{bos}
& \quad L_k= \int dx \frac{1}{4\pi}[\partial_x\phi_I^k(\partial_t-v_F\partial_x)\phi^k_I\nonumber\\ & +\partial_x\phi^k_O(-\partial_t-v_F\partial_x)\phi^k_O -\frac{v_F\lambda}{\pi}(\partial_x\phi^k_I-\partial_x\phi^k_O)^2];
\end{align}
\begin{equation}
\label{tun}
T_k=\sum_n \tilde U_n (F_O^{k\dagger}F_I^k)^n \exp(in[\phi^k_I(x=0)-\phi^k_O(x=0)])+h.c.,
\end{equation}
where $\tilde U_n=U_n\exp(i\alpha_n)$, with real $U_n$ and $\alpha_n$, are of the order of the Fourier components of the asymmetric potential, $k_F\int \exp(i2nk_Fx)U(x)dx$.
 Note that
$\alpha_n$ can be nonzero even for a symmetric potential $U(x)$
in contrast to the situation considered in Refs. \onlinecite{FSV,BFM}.
For example,
for $U\sim \delta(x-x_0)$, $\alpha_1=2k_Fx_0$.
 The above expression for the backscattering operators $T_k$ includes multi-particle backscattering processes.\cite{FSV,BFM} Such multi-particle contributions to the action
are inevitably generated under the action of the renormalization group by the interplay of a short-range Coulomb interaction and impurity potential
(see Ref.~\onlinecite{BFM} for a discussion). All prefactors
$U_n$ have the same order of magnitude and are proportional to $U(x)$. We assume
that backscattering amplitudes $U_n$ have the same order of magnitude in all three wires.
The use of the fields $\phi$ at $x=0$
in the backscattering operators
is justified, if the distance from the impurities to the junction is lower than the scale $\hbar v_F/(eV)$ set by the voltage bias.
We assume that $U_n$  are sufficiently small so that a perturbative expansion in powers of $U_n$ can be developed for the calculation of the current.
The conditions on $U_n$ will be formulated below.

Following the standard notation conventions, the Hamiltonian, corresponding to Eq.~(\ref{bos}),
can be written as

\begin{equation}
\label{ham}
H_k=\frac{v}{8\pi}\int dx[g(\partial_x\phi^k_I+\partial_x\phi^k_O)^2+\frac{1}{g}(\partial_x\phi_I^k-\partial_x\phi_O^k)^2],
\end{equation}
where the dimensionless interaction strength $g=1/\sqrt{1+2\lambda/\pi}<1$ and $v=v_F/g$.

The above action alone is not enough to describe the system and matching conditions at the junction are necessary.
As we will see, matching conditions are subject to several restrictions.
The most general matching condition has the form $F(\phi^k_{I,O}(x=0),\partial_x\phi^k_{I,O}(x=0),\partial^2_x\phi^k_{I,O},\dots)=0$, where $F$
is an arbitrary function. Below we will focus on the situation in which the system
is close to a fixed point. Then only most relevant operators should be kept in the matching conditions and hence all derivatives of the Bose-fields $\phi^k_{I, O}$
can be neglected. Next, we note that the shift of any field $\phi^k_{I,O}$ by a constant $\phi(x)\rightarrow\phi(x)+C$ is a gauge transformation that does not change the physics of the system. Hence, the matching conditions must be invariant with respect to such gauge transformations. This singles out the boundary conditions of the form
\cite{13}
\begin{equation}
\label{cond}
\phi_{O}^k=\sum_j M_{kj}\phi_I^j,
\end{equation}
where $M$ is a matrix with real matrix elements. With this matching condition, it is easy to solve the equations of motion for the chiral fields $\phi^k_O$ in the absence of the interaction ($\lambda=0$) and impurities, for arbitrary initial conditions for fields $\phi^k_{I,O}$. Substituting the solution into the commutation relations (\ref{comm}) for in- and outgoing
fields, one finds that they are compatible, provided that the matrix $M$ is orthogonal. The same condition for the matching matrix must hold for a general case
with an arbitrary interaction strength \cite{13}
since the problem can be reduced to a model of non-interacting bosons by diagonalizing
Eq. (\ref{ham}) in each wire (see the next section).

Thus, we established that $M$ is a real orthogonal matrix. We next demonstrate that the sum of the elements of each of its rows and columns equals one.\cite{13} It is again convenient
to consider the situation without electron interaction and tunneling. In that case the currents of in- and outgoing electrons in each wire are
$I^k_{I/O}=\mp v_F\rho^k_{I/O}=\pm\partial_t\phi^k_{I/O}/2\pi$. Charge conservation implies that the sum of the currents in all wires is zero at $x=0$. Taking into account
Eq.~(\ref{cond}) one finds that $\sum_k M_{kj}=1$. Multiplying Eq.~(\ref{cond}) by $M^{-1}=M^T$ and repeating the same argument one also finds that $\sum_j M_{kj}=1$. Again,
electron interactions do not affect the above result \cite{13}.

The origin of the matching conditions can be understood if one notes that they can be imposed by adding to the Hamiltonian a term of the form

\begin{equation}
\label{tun-matching}
-A\sum_k\cos(n[\phi_O^k(x=0)-\sum_j M_{kj}\phi^j_I(x=0)]),
\end{equation}
where $A$ is a large constant.
Each cosine tends to keep its argument at zero. This can be achieved simultaneously
for each cosine only if their arguments commute. This happens for a unitary $M$.
The cosines describe tunneling between
different wires. Thus, charge conservation implies $\sum_j M_{kj}=1$. This provides an alternative derivation of the matching conditions.

Since we plan to separately investigate time-reversal invariant and non-invariant systems, we next need to determine what matching conditions satisfy
the time-reversal symmetry.
This is easy as the time-reversal transformation corresponds to the change of variables $\phi^k_{I/O}\rightarrow\phi^k_{O/I}$. From Eq.~(\ref{cond}) we then see that in time-reversal-invariant systems $M=M^{-1}$. Taking into account that $M$ is orthogonal we conclude that in time-reversal invariant systems it is also symmetric.\cite{17}
The action (\ref{L}) is always time-reversal invariant. Thus, the behavior of the whole junction with respect to time reversal is determined solely by the symmetry of the matching
matrix $M$.

At this point we are in the position to give a full classification of fixed-point matching matrices.\cite{17} For time-reversal invariant systems, it is convenient to use a parametrization from Ref.~\onlinecite{16}. We discover three possibilities with time reversal symmetry:
\begin{equation}
\label{identity}
M = \left(
\begin{array}{ccc}
1& 0& 0\\
0& 1& 0\\
0& 0& 1
\end{array}\right);
\end{equation}

\begin{equation}
\label{122}
M = \frac{1}{3}\left(
\begin{array}{ccc}
-1& 2& 2\\
2& -1& 2\\
2& 2& -1
\end{array}\right);
\end{equation}

\begin{equation}
\label{time-reverse}
 M =\left(
\begin{array}{ccc}
b& a & c\\
a & c & b\\
c & b & a
\end{array}\right),
\end{equation}
where $b=\alpha(\alpha+1)/(1+\alpha+\alpha^2)$, $a=(\alpha+1)/(1+\alpha+\alpha^2)$ and $c=-\alpha/(1+\alpha+\alpha^2)$.
Eq.~(\ref{identity}) corresponds to three disconnected wires.
The cases of $\alpha=0,-1,\infty$ correspond to a junction of two wires and a detached
third wire. The physics is the same as for one linear wire and will not be discussed below.

In the absence of time-reversal symmetry we use the parametrization
from Ref.~\onlinecite{13}:
\begin{equation}
\label{no-time}
M=\left(
\begin{array}{ccc}
a& b& c\\
c& a& b\\
b& c& a
\end{array}\right),
\end{equation}
where $a=(1+2\cos\theta)/3$, $b(c)=(1-\cos\theta\pm\sqrt{3}\sin\theta)/3$
and $\theta\ne 0,\pi$. The case of $b=1$, $a=c=0$ corresponds to the island setup, depicted in Fig. 1. Note that the matching matrix (\ref{122}) has the form (\ref{no-time}) with $\theta=\pi$ so we will consider it with the case of no time-reversal symmetry.

In general, matching matrices contain negative matrix elements. This means that the incoming current in one wire may suppress outgoing current in that or other wires.
Such situation is possible in the presence of Andreev scattering, if, for example, a part of the junction is superconducting.
Negative matrix elements were also predicted in Y-shaped beam splitters for cold bosonic atoms
\cite{demler}.
Note that for noninteracting electrons no negative matrix elements are allowed \cite{aristov} since in the absence of interaction the elements of the matrix $M$ reduce to the squares of the absolute values of the elements of the scattering matrix
(cf. Sec. II).

\section{Electric current}

In order to calculate the current we need to include a voltage bias. We model Fermi-liquid leads by assuming that electron interaction in the wires is zero at
large distances $x$ from the junction.\cite{leads1,leads2,leads3} We will assume that leads 2 and 3 are kept at a zero voltage. The bias voltage $\pm V$ is applied to lead 1.
The results do not change in a more general model with the potential $\pm V$ applied to lead 1, the potential $\pm\gamma V$, $\gamma\sim 1$, applied to lead 2,
and a zero potential applied to lead 3. We will use the language of zero $\gamma$
since it is simpler. Our approach can be easily extended to a general $\gamma$.

The calculations will be based on a renormalization group approach with the voltage $V$
playing the role of the infrared cutoff. Thus, we will assume that the temperature $T<V$.
We would like to emphasize that the role of the voltage does not reduce solely to that of a cut-off; otherwise the density-driven rectification would be lost (see a discussion after Eq.
(\ref{int-repr})). It is well known that in Luttinger liquids a voltage bias can play a more prominent role than just a cut-off (see, e.g., Refs. \onlinecite{add3,add4,add5,add6}).
At the same time, the leading contribution to the total current at zero temperature can be estimated by simply  setting the renormalization group cutoff to the value of the voltage \cite{KF}.
Below we find that the rectification current is comparable with the total current in certain regimes in the absence of the time-reversal symmetry. This certainly means that in those regimes the rectification current can be computed by assuming that the voltage plays the role of a cutoff only (we however do not make such an assumption).

For a general $\gamma$, we do not expect the current to depend significantly on the temperature at $T<V$.
Such dependence can emerge at particular values of $\gamma$. In particular,
in the island setup (Fig. 1), this happens at $\gamma=0,1$.
 This can be seen from the
Keldysh perturbation theory \cite{KF}. The infrared cutoff in some integrals in the
perturbation expansion is set by ${\rm max}(T,\gamma V)$
and in some others by ${\rm max}(T,[1-\gamma]V)$.
Hence, at $\gamma=0,1$ our approach applies only for $V\sim T$ in the island setup.

We want to find the current in the first wire. At low frequencies  the current conserves and hence it is sufficient to
find the current in lead 1. It is given by the sum of the chiral incoming and outgoing currents $\pm v_F\rho_{I/O}$. The incoming current can be found from the Landauer
formula and is linear in voltage. It does not contribute to the rectification current. We thus focus on the outgoing current in lead 1. It can be found
with a generalization of the approach of Ref.~\onlinecite{FG}. Our approach is also related to that of Ref. \onlinecite{add7}.

In what follows we set the temperature $T=0$ to simplify notations. Our method can be easily generalized to finite temperatures.

Let us introduce an auxiliary field $\tilde\phi_1(x)$:
\begin{equation}
\label{aux_f}
\tilde\phi_1(x)=\phi^1_O(x), x>0;\tilde\phi_1(x)=\sum_k M_{1k}\phi^k_I(-x),x<0.
\end{equation}
The field $\tilde\phi_1$ satisfies a simple matching condition $\tilde\phi_1(+0)=\tilde\phi_1(-0)$.
Thus, the auxiliary field is a chiral field propagating through the junction.
In the stationary regime the average time-derivative of any operator is zero.
Let us now consider the operator $\hat O=\int_{-a}^{a} dx \partial_x \tilde\phi_1/2\pi$, where the integration extends between points taken in the noninteracting leads.
Its meaning is the charge carried by the mode $\tilde\phi_1$.
The equation $0=\langle d\hat O/dt\rangle=i\langle[H,\hat O]\rangle$, where the Hamiltonian $H=\sum_k(H_k+T_k)$ describes the whole system including the leads, wires and scatterers, reduces to the following relation:
\begin{equation}
\label{gen_cur}
v_F\langle\rho^1_O(a)\rangle=\langle v_F\sum_k M_{1k}\rho^k_I(a)\rangle+i\langle[\sum_k T_k,\hat O]\rangle.
\end{equation}
The left hand side is the outgoing current we want to find. The first term in the right hand side is linear in the voltage bias and cannot contribute to the rectification current.
Thus, we have to calculate only the second contribution to the right hand side.
In other words, our problem reduces to the calculation of the average backscattering  current whose operator equals $I=\sum I_l$, where
\begin{align}
\label{backscattering}
I_l = &i[ U_l (F^{k\dagger}_OF^k_I)^n \exp(i\alpha_l+in\{\phi^k_I(0)-\phi^k_O(0)\})+h.c.,\hat O] \nonumber\\
= &i[U_l(F^{k\dagger}_OF^k_I)^n\exp(i\alpha_l+in\{\sum_p (M^{-1})_{kp}\phi^p_O(0)
-\phi^k_O(0)\})
\nonumber \\ & +h.c.,\hat O] \nonumber \\
 = &i n U_l (F^{k\dagger}_OF^k_I)^n
\exp(i\alpha_l+in[\sum_p (M^{-1})_{kp}\phi^p_O(0)-\phi^k_O(0)]) \nonumber\\ &
\times [ (M^{-1})_{k1}-\delta_{k1}]+ h.c. \, .
\end{align}
Note that for any $l$ the current operator expresses via {\it all} backscattering operators in all three wires.

In order to find the average current $\langle I_l\rangle$,
we will assume that the backscattering operators $T_k$ are absent at $t=-\infty$ and are then gradually turned on.
Without the operators $T_k$ the system can be viewed as an equilibrium one in the ground state of an appropriate effective Hamiltonian.
In order to find it we introduce another auxiliary field with the structure,
similar to $\tilde\phi_1$:
\begin{equation}
\label{aux_f_2}
\bar\phi_1(x)=\phi^1_I(-x), x<0;\bar\phi_1(x)=\sum_k (\hat M^{-1})_{1k}\phi^k_O(x), x>0.
\end{equation}
In the absence of the backscattering operators $T_k$, the operator $\hat A=\int_{-\infty}^{\infty} dx\partial_x\bar\phi_1/2\pi$ commutes with the Hamiltonian.
It can be understood as the charge of a chiral mode propagating through the junction. In other words, it is an additive integral of motion. Hence, the system
can be described
by a Gibbs distribution with an appropriate thermodynamic potential conjugated to $\hat A$. The physical meaning of that thermodynamic potential is
the applied voltage bias $V$. At zero temperature, one finds that the system
is in the ground state of the effective Hamiltonian $H'= H-V\hat A$, where $H$ is the actual Hamiltonian of the Y-junction.

The current, i.e., the average of the sum of the operators $i[T_k,\hat O]$, can be now calculated with the Keldysh technique \cite{keldysh}.
It is convenient to apply the interaction representation $H\rightarrow H-V\hat A$. The interaction representation introduces time-dependence
into the operators $T_k$ and $I_l$ according to the rule:
\begin{align}
\label{int-repr}
& \exp(in[\sum_p (M^{-1})_{kp}\phi^p_O(0)-\phi^k_O(0)])\rightarrow  \nonumber\\
& \exp(iV\hat At)\exp(in[\phi^k_I(0)-\sum_p M_{kp}\phi^p_I(0)])\exp(-iV\hat At) \nonumber\\
& = \exp(in[\sum_p (M^{-1})_{kp}\phi^p_O(0)-\phi^k_O(0)+Vt(M_{k1}-\delta_{k1})]).
\end{align}
After such time-dependence is added into all backscattering operators, the contribution of the form $-V\hat A$ can be removed from the action by a linear shift
of all fields $\phi_{O/I}^k$. This does not mean, however, that the voltage would only enter the action through the time-dependence of $T_k$. Indeed, since the charge density
depends on the voltage, the amplitudes $U_k$ of the operators $T_k$ may get corrections proportional to the small voltage bias. This effect is discussed in Ref.
\onlinecite{BFM}.

The current can now be found with a perturbative expansion in powers of $U_k$ from the standard expression
\begin{equation}
\label{Keldysh}
I_l=\langle 0|S(-\infty,0) \hat I_l S(0,-\infty)|0\rangle,
\end{equation}
where $S(t,t')$ is the evolution operator from time $t'$ to time $t$ and $|0\rangle$ is the ground state of the effective Hamiltonian $H-V\hat A$.
To complete the calculation, we need the Green functions determined by the quadratic part of the action
(\ref{Lk}) and the matching conditions (\ref{cond}). They have been found in Ref.~\onlinecite{13}. A Bogoliubov transformation of the form
$\phi_{O/I}=(1/2\sqrt{g})[(1+g)\tilde\phi_{O/I}+(1-g)\tilde\phi_{I/O}]$
in each wire
allows one to obtain free chiral fields $\tilde\phi_{I/O}$ with correct
Luttinger liquid commutation relations.\cite{13} The correlation function of the incoming fields at $x=0$ is given by the free particle relation
\begin{equation}
\label{correl}
\langle\tilde\phi^k_I(t)\tilde\phi^p_I(t')\rangle=-\delta_{kp}\ln(i[t-t']/\tau_c+\delta),
\end{equation}
where $\tau_c$ is the ultraviolet cutoff time of the order of the inverse bandwidth, $\delta$ is infinitesimal.
The matching conditions for the new fields have the form \cite{13}
\begin{equation}
\label{new-match}
\tilde\phi^k_O(x=0)=\sum_p\tilde M_{kp}\tilde\phi^p_I(x=0);
\end{equation}
\begin{equation}
\label{tildeM}
\tilde M=[(1+g)-(1-g)M]^{-1}[(1+g)M-(1-g)].
\end{equation}
In the time-reversal invariant case $\tilde M=M$.
In the absence of the time-reversal symmetry, $\tilde M$ has the same general structure (\ref{no-time}) as $M$ and satisfies the same set of
constraints but the matrix elements are different:
\begin{equation}
\label{tildea}
\tilde a=\frac{3g^2-1+(3g^2+1)\cos\theta}{3(1+g^2+(g^2-1)\cos\theta)};
\end{equation}
\begin{equation}
\label{tildebc}
\tilde b(\tilde c)=\frac{2(1-\cos\theta\pm\sqrt{3} g\sin\theta)}{3(1+g^2+(g^2-1)\cos\theta)}.
\end{equation}

At this point one can write an expression for the current as an expansion in powers of backscattering amplitudes $U_l$.
Evaluation of the terms of that expansion is technically difficult and not very informative. Indeed, neither the amplitudes nor
the ultra-violet cutoff are known exactly. As a result, it is only possible to estimate the order of magnitude of each contribution.
At the same time, such estimation can be performed without explicitly calculating integrals over the Keldysh contour and
will be sufficient to find the leading power dependence of the rectification current on the bias voltage at small $U_l$
and low voltages near each of the fixed points. Such estimation will be the focus of the
remaining sections.

\section{Scaling dimensions}

It is convenient to use a renormalization group point of view for the calculation of the current.\cite{KF} In the renormalization group procedure,
the coefficient $U_l$
in the Hamiltonian and the operators $I_l$ scale as $E^{z_l-1}$, where $E$ is the energy scale and $z_l$ the scaling dimension. At the energy scale $E\sim V$
the renormalization group procedure stops. Different perturbative contributions to the current can be estimated from the scattering
theory.\cite{FSV,BFM} The current can
be expressed as an infinite sum of the contributions, proportional to the products of different combinations of $U_l$:
\begin{equation}
\label{current-scale}
I\sim V\sum {\rm const}\times \prod_l (U_l V^{z_l-1}).
\end{equation}
Strictly speaking, it is not enough to include only backscattering operators from the bare tunneling Hamiltonian (\ref{tun}) in the expansion (\ref{current-scale}).
All operators generated by the renormalization group must also be included. Their general structure is
\begin{align}
\label{gen}
\hat W_l= & W_l\exp(i\alpha_l+i\sum_k n_k[\phi^k_O(x=0)-\phi^k_I(x=0)]) \nonumber\\
= & W_l\exp(i\alpha_l+i\sqrt{g}\sum_k\sum_p n_k[\tilde M_{kp}-\delta_{kp}]\tilde\phi^p_I).
\end{align}
For simplicity we use the same notation $\alpha_l$ for the phases of $W_l$ and $U_k$.
Note that it is always sufficient to keep only two different values of $k$ in the sum in Eq.~(\ref{gen}), since $\sum_{k=1}^3[\tilde M_{kp}-\delta_{kp}]=0$.
This means that each possible operator $\hat W_p$ can be generated from a product of two operators $T_k$ (\ref{tun}) and hence $W_p\sim U_l^2$. Since Klein factors
do not change the scaling dimensions of $W_l$, we omit them in Eq.~(\ref{gen}).

To complete the calculation of the current we need to find $z_l$. A straightforward calculation based on Eq.~(\ref{correl}) yields for the operator (\ref{gen})
with an arbitrary choice of $n_k$:
\begin{equation}
\label{z-exp}
z_l=g\left(\sum (n_i)^2-\sum n_i n_j \tilde M_{ij}\right).
\end{equation}
In the absence of the time-reversal symmetry the above expression greatly simplifies:
\begin{equation}
\label{z-exp-no-time}
z_l=g(1-\tilde a)[n_1^2+n_2^2+n_3^2-n_1n_2-n_2n_3-n_3n_1].
\end{equation}
In the low-voltage regime, the main contribution comes from the operators with the smallest $z_l$.
One easily sees that it equals
\begin{equation}
\label{z-min-no-time}
z_{\rm min}=g(1-\tilde a)
\end{equation}
and is achieved, if two of the coefficients $n_i=0$ with the remaining one being $\pm 1$.

In the time-reversal invariant system, the expression for the scaling dimensions is more complicated. It can be simplified by setting $m=n_1-n_2$
and $k=n_2-n_3$. Using the matching matrix (\ref{time-reverse}), one finds:
\begin{equation}
\label{z-exp-time}
z=g\frac{[k(b+c)+mc]^2}{b+c}=\frac{g}{1+\alpha+\alpha^2}(k\alpha-m)^2.
\end{equation}
We will assume for simplicity that $\alpha$ is irrational. We will thus avoid situations
with $z=0$ at some $k$ and $m$. Certainly, such situations correspond to the operators $W_l$
with no $\phi$-dependence. Such operators cannot affect transport.

The renormalization group procedure only applies if all $U_l$ are small at the initial energy scale $\sim 1/\tau_c$ and remain small up to the scale $V$.
In the case without time-reversal symmetry, this condition can be easily expressed in terms of $z_{\rm min}$:
\begin{equation}
\label{U-max}
U_l\tau_c<{\rm const} (V\tau_c)^{1-z_{\rm min}}={\rm const}V^{1-g(1-\tilde a)}.
\end{equation}
The above equation assumes that $1-z_{\rm min}>0$, i.e., $U_l$ is relevant. If $1-z_{\rm min}<0$ then the only restriction on $U_l$ is $U_l\tau_c<1$.
What happens in the presence of the time-reversal symmetry will be addressed in subsequent sections.

\section{Rectification current}

The above discussion applies to all contributions to the current, both even and odd in the voltage bias. We are interested in the even contribution, i.e., the rectification current $I_r(V)=[I(V)+I(-V)]/2$.
What terms of the perturbation expansion contribute to the rectification current? The answer to this question can be obtained from symmetry considerations.

In what follows we will need the commutation relations for the Klein factors in (\ref{tun}). Using the commutation relations
$\psi_O^{k\dagger}(0)\psi_I^k(\epsilon)=
-\psi^k_I(\epsilon)\psi_O^{k\dagger}(0)$ and
$[\phi_I^k(x),\phi_I^k(y)]=-i\pi {\rm sign}(x-y) $, an expression for $\psi_O^k$ in terms of $\phi_I^k$ and the Baker-Hausdorff
formula, one finds
\begin{equation}
\label{klein-comm}
F_O^{k\dagger} F_I^k=\exp(i\pi[1-M_{kk}])F_I^k F_O^{k\dagger}.
\end{equation}

Next, some terms of the perturbative expansion (\ref{current-scale}) are zero identically as only certain combinations of the vertex operators in $T_k$ and $I_l$
produce non-zero results after averaging with respect to the quadratic part of the action (\ref{bos}). The condition is well known:
\begin{equation}
\label{nonzero}
\langle \prod_l\exp(i\sum_k c_{lk}\phi_I^k(0))\rangle\ne 0
{\rm~~only~if~for~each~} k~ \sum_l c_{lk}=0.
\end{equation}

Let us now apply the above results to possible contributions of different orders of the perturbation theory to the rectification current.
Strong limitations emerge for second order contributions. Indeed, Eq.~(\ref{nonzero}) implies that any second order contribution, proportional to an operator
$U_l$, must also contain its Hermitian conjugated operator $U^\dagger_l$. Hence, the phase factors $\exp(\pm i\alpha_l)$ cancel each other and drop out from the
expression for the current. Thus, we can just set $\alpha_l=0$. Let us now compare the currents at the bias voltages $V$ and $-V$. In the case with the bias $-V$,
we
make the change of variables $\phi^k_{I/O}\rightarrow-\phi^k_{I/O}$. This transformation does not affect the form of the quadratic part of the action (\ref{bos}) and the linear matching
conditions. Let us also change the order of the Klein factors in the tunneling and current operators using the commutation relation (\ref{klein-comm})
and redefine $F\rightarrow F^\dagger, F^\dagger\rightarrow F$.
Since $U_l$ pairs up with $U_l^\dagger$ in the nonzero
second-order
contribution to the current, one can easily see that the phase factors
$\exp(\pm i\pi[1-M_{kk}])$
in the commutation relation (\ref{klein-comm})
drop out. At the same time, if we omit the phase factors $\exp(\pm i\alpha_l)$ and $\exp(\pm i\pi[1-M_{kk}])$ from the action and current operators,
we discover that the action assumes precisely the same form as for the voltage bias $V$. On the other hand, the current operator $I_l$ (\ref{backscattering})
changes its sign. This means that the average of $I_l$ also changes its sign and hence the second order term
of the order $U_l^2 V^{2z_l-1}$
does not contribute to the rectification current.

The above argument assumes that the backscattering amplitudes $U_l$ are independent of the voltage bias. Since the injected charge density depends on the voltage in our setup, $U_l$ can exhibit a weak linear dependence on the bias voltage. This means that second order terms can in fact lead to a ``density-driven'' rectification effect
but its amplitude is suppressed by an additional factor $V\tau_c$,
where the ultra-violet cutoff $\tau_c$ is of the order of the inverse band width.
Thus, the second order contribution to the ratchet current scales as
\begin{equation}
\label{2order}
I^{(2)}\sim U_l^2 V^{2z_l}.
\end{equation}

Beyond the second order, no additional general restrictions on possible contributions to the rectification current can be derived. Particular contributions to
the ratchet current can disappear at particular values of the phases $\alpha_l$
and interaction strength $g$.

\section{Three weakly connected wires}

\begin{figure}
\centering
\includegraphics[width=2in]{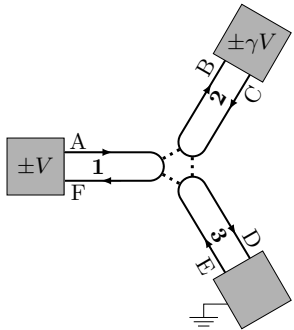}
\caption{Schematic picture of a Y-junction of three weakly connected wires. Voltage bias $\pm V$ is applied to wire 1, and bias $\pm \gamma V$ to wire 2. Wire 3
is connected to the ground. We calculate the dc current in wire 1.}\label{fig4}
\end{figure}

So far we discussed general features of Y-junctions. We now determine the rectification current in different setups. As a warm-up exercise we investigate
the simplest situation of three almost disconnected wires, Fig. 4. This situation corresponds to the matching matrix (\ref{identity}). The rectification current is much weaker in this limit than for other matching conditions. At the same time, the qualitative picture is quite similar.

There is no current at all in our original model with the action (\ref{L}) and backscattering operators (\ref{Tk}) as it describes three disconnected conductors.
We thus modify the model in this section. Instead of the backscattering operators (\ref{Tk}) we consider weak tunneling between the wires. The tunneling operators
have the form
\begin{align}
\label{3-tun}
T=&\sum_{k=1}^3 T_k;\nonumber\\ T_k=&F_{k+1}^\dagger F_kU_ke^{i\alpha_k}\exp(i\phi_k(0)-i\phi_{k+1}(0))+h.c.,
\end{align}
where $\phi_k(0)=\phi_I^k(0)=\phi_O^k(0)$, $F_k$ are Klein factors and we use the convention $3+1\equiv 1$.
The commutation relations for the Klein factors are $F_k F_q=-F_q F_k, k\ne q$.
We assume that wires 3 is kept at zero voltage. The chemical potential of the first wire is $\pm V$. The second wire experiences the bias $\pm\gamma V$. Note a different meaning of the amplitudes $U_k$ and phases $\alpha_k$ from the model
with three weak scatterers discussed above. In Eq. (\ref{3-tun}),
the phases $\alpha_k$ describe the Aharonov-Bohm effect due to the magnetic field through the junction. On the other hand, in the model with weak scatterers, all information about the magnetic flux
is contained in the matching conditions.
The voltage bias can be included in the action in the form of the time-dependence of the tunneling operators, similar to the discussion above.
We will not need explicit expressions below.
The operator of the current, tunneling between wire 1 and wires 2 and 3, is
\begin{equation}
\label{3-current-op}
I=iT_1-iT_1^\dagger-iT_3+iT_3^\dagger.
\end{equation}

The current can be estimated from a renormalization group procedure that stops at the scale $E\sim V$. At that scale
$U_k\rightarrow U_k (V\tau_c)^{1/g-1}$, where $g<1$ characterizes the interaction strength. The second order contribution to the
rectification current can now be easily found and scales as
\begin{equation}
\label{3-I-2}
I^{(2)}\sim U^2 V^{2/g}\tau_c^{2/g+1},
\end{equation}
where $U\sim U_l\ll 1/\tau_c$.

The $n$-th order contribution to the current cannot exceed $\sim (U\tau_c)^n (V\tau_c)^{n/g-n+1}/\tau_c$, $n\geq3$. Thus, the leading higher-order contribution corresponds to $n=3$,
\begin{equation}
\label{3-I-3}
I^{(3)}\sim U^3 V^{3/g-2}\tau_c^{3/g}.
\end{equation}
Interestingly, at low voltages and $g>1/2$, the third order contribution exceeds the second order contribution.

The above discussion ignored the issue of the time-reversal symmetry.
Depending on the presence or absence of the magnetic field through the junction, the system can have or have no time-reversal symmetry.\cite{4,5}
It is instructive to investigate the effect of the symmetry breaking.
In the presence of the time-reversal symmetry the magnetic flux is zero and all $\alpha_k=0$. Indeed, the time-reversal transformation can be represented
as $\phi_I^k\rightarrow\phi_O^k,\phi^k_O\rightarrow\phi_I^k, \alpha_k\rightarrow -\alpha_k$. Since $\phi_I^k(0)=\phi_O^k(0)$, we conclude
that $\alpha_k=-\alpha_k=0$ at zero magnetic field.
On the other hand, in the presence of the magnetic field $\alpha_k\ne 0$.

The second order contribution to the rectification current does not depend on the phases $\alpha_k$.
Let us compare the third order contributions for different values of $\alpha_k$.
Nonzero third order contributions to the current
originate only from the product of all three operators $T_k$ or all three operators $T_k^\dagger$ in the perturbative expansion of the Keldysh expression for the current (\ref{Keldysh}). The first contribution, $I^{(3)}_1$, is proportional to $\exp(i\sum_k\alpha_k)$,
$I^{(3)}_1(V)=\exp(i\sum_k\alpha_k) J_1$, and the second contribution is proportional to $\exp(-i\sum_k\alpha_k)$, $I^{(3)}_2(V)=\exp(-i\sum_k\alpha_k) J_2$.
Let us change the voltage sign. Simultaneously, we change $\phi_{O/I}^k\rightarrow -\phi_{O/I}^k$,
change the order of Klein factors in all operators and redefine $F_k\rightarrow F_k^\dagger, F_k^\dagger\rightarrow F_k$.
This transformation is equivalent to changing the sign of $\alpha_k$ in all operators and  simultaneously changing the overall sign of the tunneling Hamiltonian (\ref{3-tun})
but not the sign of the current operator. Hence,
the third order current $I^{(3)}_1(-V)+I^{(3)}_2(-V)=\exp(-i\sum_k\alpha_k) J_1+\exp(i\sum_k\alpha_k) J_2$. We see that the current is an even function of the voltage
in the presence of the time-reversal symmetry, i.e., at $\sum_k \alpha_k=0$. The third order current is an odd function of the voltage and does not contribute to the rectification effect, if $\sum_k\alpha_k=\pm \pi/2$. Thus, for three weakly connected wires, magnetic field suppresses rectification. We will see below that typically it has an opposite effect and rectification is stronger in the absence of the time-reversal symmetry.
Note that in the above example, the ratchet current returns to its maximal value at
$\sum_k\alpha_k=\pi$.

Let us finally compare our results with the case of a linear wire with a high asymmetric potential barrier or, equivalently, two weakly connected wires. In
that case the rectification current is given \cite{FSV} by Eq.~(\ref{3-I-2}). Thus, at $g>1/2$ the rectification effect is stronger in a Y-junction than in a linear
wire with the same interaction strength.

\section{Rectification effect in the absence of the time-reversal symmetry} \label{sec:no-time}

Here we address matching matrices of the form (\ref{no-time}).
In particular, the results of this section are relevant for the island setup, Fig. 1.
Our analysis also applies to the
time-reversal-invariant situation with the matching matrix (\ref{122}).
Below we find that at sufficiently low voltages the rectification current scales as
Eq. (\ref{3-no-time}) with $U$ being the impurity potential strength,  $V$ the voltage,
$g<1$ the Luttinger liquid parameter, characterizing electrostatic interactions, $\tau_c$ the ultraviolet cutoff of the order of the inverse band width, and $\tilde a$ is determined by the matching conditions according to Eq. (\ref{tildea}).

Similar to the previous section we need to compare the leading second and third order contributions to the rectification current.
They are determined by the operators of the form $T_k\sim\exp(i[\phi_I^k-\phi_O^k])$ with the  minimal scaling dimension (\ref{z-min-no-time}).
The leading second order contribution scales as
$I^{(2)}\sim (U\tau_c)^2 (V\tau_c)^{2g(1-\tilde a)}/\tau_c$, where $\tilde a$ is given by Eq.~(\ref{tildea}) and satisfies $-1/3 \le\tilde a\le 1$.
The leading third order contribution comes from the terms in the perturbation expansion proportional to the product of all three $T_k$
(or all three $T_k^\dagger$). It scales as

\begin{equation}
\label{3-no-time}
I^{(3)}\sim U^3 V^{3g(1-\tilde a)-2}\tau_c^{3g(1-\tilde a)}.
\end{equation}

At small $U$ the second order contribution always exceeds higher-order contributions. Interestingly, however, at greater $U$ which are still within the
region of validity of the perturbation expansion, the third order contribution is leading.

Let us show that the third order contribution dominates.
We will focus on the largest $U_l$ accessible with the perturbation expansion: $U_l\sim 1/\tau_c$ at $g(1-\tilde a)>1$ when $U_l$ is irrelevant; and $U_l\sim (V\tau_c)^{1-g(1-\tilde a)}/\tau_c$ at $g(1-\tilde a)<1$ when $U_l$ is relevant. In all cases the electron interaction
is repulsive, i.e., $g<1$.
Let us first consider the case of  $g(1-\tilde a)>1$. According to Eq.~(\ref{tildea}),
$\tilde a\ge -1/3$. Hence $g(1-\tilde a)<4/3$. The ratio of the second order contribution to the third order contribution scales as $(U\tau_c)^{-1}(V\tau_c)^{2-g(1-\tilde a)}
\sim (V\tau_c)^{2-g(1-\tilde a)}<1$. Thus, the third order contribution dominates indeed. The case of $g(1-\tilde a)<1$ is also easy. In the limit of
$U_l\sim (V\tau_c)^{1-g(1-\tilde a)}/\tau_c$ we find $\tau_c I^{(2)}\sim (V\tau_c)^2$ and
$I^{(3)}\sim V\gg I^{(2)}$. Interestingly, the third order current may become comparable
to the total ac current $\sim e^2V/h$ through the junction at $g(1-\tilde a)<1$.
Note also that the exponent  $3g(1-\tilde a)-2$ in the voltage dependence of the dc current
(\ref{3-no-time}) is negative at $g(1-\tilde a)<2/3$. This corresponds to the dc current
{\it increase} as the ac voltage decreases.

The above calculation applies for the voltage interval $1/\tau_c>V$ at $g(1-\tilde a)>1$  and $1/\tau_c>V>(U\tau_c)^{1/[1-g(1-\tilde a)]}/\tau_c$
at $g(1-\tilde a)<1$. The left inequality is dictated by the applicability of the
Luttinger liquid model. The right inequality is determined by the validity of the perturbation theory. One cannot calculate the rectification current
at lower voltages with the above perturbative approach.
However, it is obvious that $I=0$ at $V=0$. From this we conclude that there is a bump on the voltage dependence of the rectification current at
$g(1-\tilde a)<2/3$:
it grows as a function of the voltage at low voltages and achieves its maximal value of the order of $e^2V/h$ at $V=V^*\sim U^{1/[1-g(1-\tilde a)]}$. In the last equation we omitted
a power of $\tau_c$ as this does not lead to a confusion.

Finally, let us compare the result with the case of a linear wire with the same interaction strength $g$ in the presence of an asymmetric impurity. In that case the maximal
rectification current, calculated in the perturbative regime in Ref. \onlinecite{FSV}, scales as $V^{1+3g}<V$ at $g<1/3$ (we omit powers of $\tau_c$). The rectification current does not exceed $V^2/\tau_c$ at $g>1/3$.
In a Y-junction, at $1-g(1-\tilde a)>0$ the rectification current $\sim V$
at $V\sim V^*$ is always greater than in a linear wire. On the other hand,
  a negative $1-g(1-\tilde a)$ implies $g>3/4$.  The maximal
rectification current corresponds then to $U\sim 1/\tau_c$ and scales as $\sim V^{3g(1-\tilde a)-2}>V^2$. Thus, the maximal ratio of the rectification and total currents is always higher in a Y-junction than in a linear wire in the Luttinger liquid regime $V\ll 1/\tau_c$.

\section{Rectification effect in the presence of the time-reversal symmetry}

This case is more complicated than the situation without the symmetry.
The summary of the results is given in section IX.C.
We will need to take into account many different backscattering operators.
Thus, it is important to classify them. The classification relies on the matching conditions at the junction.

The matching matrix is given by equation (\ref{time-reverse}) where the expressions
for $a=f_1(\alpha), b=f_2(\alpha), c=f_3(\alpha)$ in the parametrization \cite{17} are written under the equation. It is always possible to redefine $\alpha$ in such a way
that $a=f_k(\alpha), b=f_l(\alpha), c=f_n(\alpha)$, where $(k,l,n)$ is an arbitrary transposition of $(1,2,3)$. Indeed, the change of the variable $\alpha\rightarrow 1/\alpha$
exchanges $f_1$ and $f_2$; $\alpha\rightarrow -(1+\alpha)$ exchanges $f_1$ and $f_3$; $\alpha\rightarrow -\alpha/(1+\alpha)$ exchanges $f_2$ and $f_3$. Any other transposition is a superposition of the above three. Since $M$ is an orthogonal matrix, at least one of the elements must be negative or zero. Otherwise, different rows cannot be orthogonal. Without loss of generality we may assume that $c\le 0$ and hence $\alpha\ge 0$.
This can always be achieved by renumbering the wires and redefining $\alpha$.
Note that $a,b\ge 0$.
Similarly, without loss of generality we may assume that $a\ge b$. Hence, $\alpha\le 1$. Note that $a\ge 2/3$. Thus, the ranges of the parameters that we consider are: $0\leq\alpha\leq1$, $-1/3\leq c\leq0$, $0\leq b\leq2/3$ and $2/3\leq a\leq1$.

As has already been discussed, we need to deal with two types of backscattering operators in the action and the current operator:
\begin{description}
\item{$(*)$} $U_l\sim U\exp(in_k[\phi^k_O(x=0)-\phi^k_I(x=0)])$;

\item{$(**)$} $W_l\sim U^2\exp(in_1[\phi^1_O(x=0)-\phi^1_I(x=0)]+in_3[\phi^3_O(x=0)-\phi^3_I(x=0)])$.
\end{description}
Note that we do not need to include  a contribution of the form $n_2[\phi^2_O(x=0)-\phi^2_I(x=0]$ in the exponent in $W_l$.

We will first concentrate on the operators of the first type and determine the leading contribution to the ratchet current
in the absence of the operators of the form $W_l$. Next, we will check what changes after $W_l$ are taken into account. We summarize our findings in Sec. IX.C.

\subsection{Contributions from operators $U_l$}

The scaling dimensions of the operators $U_l$ are simply

\begin{equation}
\label{U-l-scale}
z_1=n_1^2g(1-b); z_2=n_2^2g(1-c); z_3=n_3^2g(1-a).
\end{equation}
The lowest scaling dimension is $z_3$ with $n_3=1$. It sets the scale of the maximal
$U$ in the perturbation theory,
$U\sim (V\tau_c)^{1-g(1-a)}/\tau_c$. Strictly speaking we need to know the scaling behavior
of the operators $W_l$ to determine the maximal allowed value of $U$ in the perturbative regime. We will see in the next subsection that the operators $W_l$
do not change the above expression for the maximal $U$. We thus immediately find the leading second-order contribution to the rectification current
\begin{equation}
\label{second-order}
I^{(2)}\sim \frac{1}{\tau_c}(U\tau_c)^2 (V\tau_c)^{2g(1-a)}.
\end{equation}

What about the third and higher orders?
It turns out that the leading contribution is third order and comes from the operators
$U\exp(2i[\phi^3_O(x=0)-\phi^3_I(x=0)])$, $U\exp(-i[\phi^3_O(x=0)-\phi^3_I(x=0)])$ and $U\exp(-i[\phi^3_O(x=0)-\phi^3_I(x=0)])$. It
scales as
\begin{equation}
\label{third-order-main}
I^{(3)}\sim \frac{1}{\tau_c}(U\tau_c)^3(V\tau_c)^{6g[1-a]-2}
\end{equation}

To see that (\ref{third-order-main}) is the leading contribution, we first
note that $I^{(3)}$, Eq.~(\ref{third-order-main}), exceeds $I^{(2)}$ at the maximal
allowed
$U\sim (V\tau_c)^{1-g(1-a)}/\tau_c$. Indeed, at such $U$, $I^{(2)}\tau_c\sim (V\tau_c)^2$
while $I^{(3)}\tau_c\sim (V\tau_c)^{1+3g[1-a]}>(V\tau_c)^{1+3g\times 1/3}>(V\tau_c)^2$. Next, let us compare the contribution (\ref{third-order-main}) with other higher order contributions
to the current,
\begin{equation}
\label{third}
I'\sim V\prod_l\left(UV^{z_l-1}\tau_c^{z_l}\right)\sim V\prod_l \left([V\tau_c]^{z_l-g[1-a]}\right).
\end{equation}
If an operator with the scaling dimension $z_l=n^2_2 g[1-c]$ enters
the above expression
then (\ref{third}) is smaller than (\ref{third-order-main}).
Indeed, any contribution to the rectification current that contains such operators
satisfies the inequality
$I'\tau_c<(V\tau_c)^{1+g[a-c]}=(V\tau_c)^{1+g[1+2\alpha]/[1+\alpha+\alpha^2]}<(V\tau_c)^{1+3g[1-a]}$. Thus, only contributions with the operators $U_l\sim U\exp(in_{1l}[\phi^1_O(x=0)-\phi^1_I(x=0)])$
and $U_l\sim U\exp(in_{3l}[\phi^3_O(x=0)-\phi^3_I(x=0)])$ have to be considered. If one of the contributing operators has $|n_{1l}|>1$ or $|n_{3l}|>1$
then it is obvious that the contribution cannot exceed (\ref{third-order-main}). We thus consider the case with all $n_{1l}, n_{3l}=\pm 1$.
A nonzero contribution to the current requires $(\sum_l n_{1l})[\phi^1_O(x=0)-\phi^1_I(x=0)]+(\sum_l n_{3l})[\phi^3_O(x=0)-\phi^3_I(x=0)]=0$.
For a general transcendental $\alpha$ this equality is satisfied only if $\sum_l n_{1l}=\sum_l n_{3l}=0$. Thus, we consider contributions in which the operator $U_1= U_1F^{1\dagger}_O F^1_I\exp(i\alpha_1+i[\phi^1_O(x=0)-\phi^1_I(x=0)])$
enters the same number of times as the operator $U_1^\dagger$ and the operator
$U_3= U_3F^{3\dagger}_OF^3_I\exp(i\alpha_3+i[\phi^3_O(x=0)-\phi^3_I(x=0)])$
enters the same number of times as the operator $U_3^\dagger$. Hence, the phases $\alpha_{1,3}$ drop out from the final answer.
Let us now change the voltage sign, perform the transformation $\phi^k_{I/O}\rightarrow -\phi^k_{I/O}$, change the order of the Klein factors
in each term and redefine $F_{I/O}^k\rightarrow F_{I/O}^{k\dagger},F_{I/O}^{k\dagger}\rightarrow F_{I/O}^k$.
Changing the order of the Klein factors introduces complex conjugate phase factors into $U_k$ and $U^\dagger_k$. Hence, we can ignore both those phase
factors and $\alpha_k$ since neither affects the final result. On the other hand, if we ignore the phase factors, we discover that the backscattering part
of the Hamiltonian does not change under our transformation while the current operators $I_l$ change their signs. This means, in turn, that the contribution to the total
current we are calculating
changes its sign when the bias voltage changes its sign.
Hence, it does not contribute to the {\it rectification} current. Thus, Eq.~(\ref{third-order-main})
describes the main contribution to the rectification current.

\subsection{Contributions from operators $W_l$}

We now find the leading contribution to the rectification current that contains operators $W_l$.
First of all, let us check that renormalized operators $W_l(V )\sim W_l(E\sim 1/\tau_c)(V\tau_c)^{z_l-1}$ remain small at $U<(V\tau_c)^{1-g[1-a]}/\tau_c$.
Indeed, $W_l(1/\tau_c)\sim U^2$ and $z_l\ge 0$. Hence, $W_l(V)\tau_c<(V\tau_c)^{1-2g[1-a]}<(V\tau_c)^{1-2g/3}<(V\tau_c)^{1/3}\ll 1$. Thus, all
renormalized operators remain small and the maximal $U$ for which the perturbation theory
can be used was found correctly in the previous subsection.
In what follows we focus on the case with $U$ of the order of its maximal allowed value.
The opposite limit $U\rightarrow 0$ is trivial since in that limit the second order contribution in $U$ always dominates the ratchet current.

Next, let us check how the operators $W_l$ affect second order contributions to the rectification current. The corresponding second order contribution
$I^{(2)}_W\sim [W_l(V)]^2V^2/\tau_c^3\le (U\tau_c)^4/\tau_c\sim (V\tau_c)^{4-4g[1-a]}/\tau_c$. We need to compare it with the current (\ref{third-order-main}), $I^{(3)}\sim (V\tau_c)^{1+3g[1-a]}/\tau_c$.
One easily sees that $4-4g[1-a]>1+3g[1-a]$ since $g[1-a]<1/3$. Thus, the second order contribution to the rectification current can be neglected
in comparison with the third order contribution even after $W_l$ are taken into account.

At the same time, in a certain region of parameters the dominant higher-order contribution to the rectification current contains an operator $W_l$.
As we will see, such contribution contains exactly one operator $W_l$. We thus begin our analysis by excluding contributions which contain three or more
$W_l$'s. Indeed, any such contribution $I^{3W}\le V (W_l/V)^3\sim (U\tau_c)^6/(V^2\tau_c)\sim (V\tau_c)^{4-6g[1-a]}/\tau_c$. We need to compare this estimate with
$I^{(3)}\sim (V\tau_c)^{1+3g[1-a]}/\tau_c$.
One easily sees that $[4-6g(1-a)]>[1+3g(1-a)]$ and hence we do not need to take into account contributions with three or more operators $W_l$.

Thus, it remains to consider higher-order contributions with one or two operators $W_l$. We first consider contributions with exactly one operator
$\hat W_l= W_l\exp(i\alpha_l+\sum_k s_k[\phi^k_O(x=0)-\phi^k_I(x=0)])$, where $s_2$ can be set to zero as discussed above.
The contribution also contains at least two operators $U_l$. It is sufficient to consider the case with all $n_k=\pm 1$ in the definition
of the operators $U_l$
$(*)$. Indeed, a contribution
with an operator with $|n_k|=p>1$ can be increased by substituting it with $p$ operators
$U_{k,\pm}=\exp(\pm i[\phi^{k}_I-\phi^{k}_{O}])$.
Moreover, we can assume that for each $k$ only operators $U_{k,+}$ or only operators $U_{k,-}$ enter.
Let us denote
the number of the operators of the form  $\exp(\pm i[\phi^k_{I}-\phi^k_{O}])$ in the expression for the contribution under consideration as $l_k$.
Then we can estimate the contribution to the rectification current as
\begin{equation}
\label{est-one-w}
I^{(3')}\sim V(U^2V^{z-1})[UV^{g(1-a)-1}]^{l_3}[UV^{g[1-b]-1}]^{l_1}[UV^{g(1-c)-1}]^{l_2},
\end{equation}
where $z=g[s_3\alpha+s_1]^2/(1+\alpha+\alpha^2)$.
The same argument as in the previous subsection shows that the maximal contribution corresponds to $l_2=0$. Then for a general $\alpha$,
the product of the vertex operators $U_l, W_l$ gives a nonzero result after averaging with respect to the quadratic part of the action only if
$l_1=|s_1|$ and $l_3=|s_3|$. Note that we can assume that both $s_1$ and $s_3$ are nonzero and $s_1\ne s_3$.
Otherwise the operator $W_l$ would have the same from as one of the operators $U_l$
and the analysis from the previous subsection would apply.

We want to compare the contributions (\ref{third-order-main}) and (\ref{est-one-w}). At $U\sim (V\tau_c)^{1-g[1-a]}/\tau_c$ we find
\begin{align}
\label{ratio-3-3}
&\log [I^{(3')}/I^{(3)}] = \gamma(s_1,s_3)\log (V\tau_c); \nonumber \\
&\gamma =1+\frac{g}{1+\alpha+\alpha^2}[-5\alpha^2+|s_1|(1-\alpha^2)+(s_1+s_3\alpha)^2].
\end{align}
If $\gamma$ is positive for every choice of $s_1,s_3,s_1-s_3\ne 0$ then $I^{(3)}$ is
the main contribution to the rectification current. If $\gamma<0$ for a certain choice of $s_1,s_3$ then the leading
contribution comes from $I^{(3')}$. We thus want to investigate at what conditions $\gamma<0$ and find what choice of $s_1$ and $s_2$
minimizes $\gamma$. That choice determines the power dependence of the rectification current (\ref{est-one-w}) on the voltage.

First, let us prove that $\gamma$ is minimal, if $|s_1|=1$. Indeed, let us compare $\gamma(s_1=1,s_3=-1)$ with $\gamma(p_1,p_2)$, where $|p_1|>1$ and
$p_2$ is arbitrary. One finds
\begin{align}
\label{vspmg}
& \gamma(p_1,p_2)-\gamma(1,-1)\nonumber\\
& =\frac{g}{1+\alpha+\alpha^2}[(|p_1|-1)(1-\alpha^2)+(p_1+\alpha p_3)^2-(1-\alpha)^2] \nonumber\\
&\ge\frac{g}{1+\alpha+\alpha^2}(1-\alpha)[(|p_1|-1)(1+\alpha)-(1-\alpha)]>0.
\end{align}
Thus, we can focus on $s_1=1$ (the case of $s_1=-1$ is completely analogous).

Next, we prove that a negative $\gamma(1,s_3)$ is minimal at $s_3=-1$.
Indeed, a negative $\gamma(1,s_3)$ implies that $1+g[1-6\alpha^2]/[1+\alpha+\alpha^2]<0$. Hence,
$\alpha>(1+\sqrt{41})/10>0.7$. Taking into account that $\alpha<1$, we find that $(s_1+s_3\alpha)^2=(1+s_3\alpha)^2$ is minimal at $s_3=-1$.
This allows one to establish that $\gamma$ is minimal at $s_3=-1$.

The remaining task is simple. We just determine at what conditions $\gamma(1,-1)$ is negative, i.e., we need to investigate the inequality
\begin{equation}
\label{inv-ineq}
\gamma(1,-1)=1+\frac{g[2-2\alpha-5\alpha^2]}{1+\alpha+\alpha^2}<0.
\end{equation}
One easily sees that $\gamma$ can only be negative, if $g>3/5$, i.e., for relatively weak repulsive electron interaction.
$\gamma$ is negative in the largest interval of $\alpha$ at $g\rightarrow 1$. In that case, $\gamma<0$ for $\alpha>3/4$. The contributions with $W_l$ matter only if $\gamma(1,-1)<0$.
They give rise to the  rectification current of the form (\ref{answer2}) at negative $\gamma$.

The last question we must address in this subsection concerns the role of the contributions to the rectification current with two operators of the form $(**)$.
An estimation of such contributions is similar to (\ref{est-one-w}):
\begin{equation}
\label{est-two-w}
I^{(3'')}\sim V (U^2/V)^2V^{z_x+z_y}[UV^{g(1-a)-1}]^{l_3}[UV^{g[1-b]-1}]^{l_1},
\end{equation}
where $z_{x,y}=g[s^{x,y}_3\alpha+s^{x,y}_1]^2/(1+\alpha+\alpha^2)$ are the scaling dimensions of the operators
$W_{x,y}\sim U^2\exp(is^{x,y}_1[\phi^1_O(x=0)-\phi^1_I(x=0)]+is^{x,y}_3[\phi^3_O(x=0)-\phi^3_I(x=0)])$.

As above, we first divide $I^{(3'')}$ by $I^{(3)}$ and ask when the ratio is greater than 1:
\begin{align}
\label{ratio-3-3-new}
&\log [I^{(3'')}/I^{(3)}]=\mu(s^x_1,s^x_3,s_1^y,s_3^y)\log (V\tau_c) ;  \nonumber \\
&\mu=2+\frac{g}{1+\alpha+\alpha^2} (-7\alpha^2+l_1(1-\alpha^2)+\sum_{r=x,y}(s_1^r+s_3^r\alpha)^2).
\end{align}
If $\mu<0$ then $I^{(3'')}>I^{(3)}$.
Obviously, a negative $\mu$ implies that $0>2-7g\alpha^2/(1+\alpha+\alpha^2)$ and hence $\alpha>(1+\sqrt{11})/5>0.8$.
Let us now compare $\mu$ and $\gamma(1,-1)$ at $\alpha>0.8$. One finds

\begin{equation}
\label{mu-gamma}
\mu-\gamma(1,-1)\ge 1+\frac{2g}{1+\alpha+\alpha^2}(\alpha-\alpha^2-1).
\end{equation}
One easily checks that the above difference is positive at $\alpha>0.8$. Hence, contributions with one or no operators $W_l$ are always more important than contributions
with two such operators.

\subsection{Summary for systems with time-reversal symmetry}

We found that the leading contribution to the rectification current depends on the sign of $\gamma(1,-1)$, Eq.~(\ref{inv-ineq}). For strong repulsive interaction ($g<3/5$), $\gamma(1,-1)$ is always positive. If $\gamma>0$, at sufficiently low voltages $V\sim V^*=(\tau_c U)^{1/[1-g(1-a)]}/\tau_c$ the current scales
as

\begin{equation}
\label{answer1}
I_r\sim U^3V^{\frac{6g\alpha^2}{1+\alpha+\alpha^2}-2}.
\end{equation}
At a negative $\gamma$ and low voltages $V\sim V^*$

\begin{equation}
\label{answer2}
I_r\sim U^4 V^{\frac{2g(\alpha^2-\alpha+1)}{1+\alpha+\alpha^2}-2}.
\end{equation}

The exponents in both voltage dependences are negative.
This is related to the fact that our calculations are only valid in an interval of low voltages, $1/\tau_c\gg V>V^*=(\tau_c U)^{1/[1-g(1-a)]}/\tau_c$.
Similar to Section VIII the $I$-$V$ curve for the rectification current exhibits a bump at $V\sim V^*$. The maximal rectification current at $\gamma>0$

\begin{equation}
\label{answer3}
I_{\rm max}\sim \frac{e^2V}{h} (V\tau_c)^{\frac{3g\alpha^2}{1+\alpha+\alpha^2}}.
\end{equation}
At $\gamma<0$

\begin{equation}
\label{answer4}
I_{\rm max}\sim \frac{e^2V}{h}(V\tau_c)^{1-2g\frac{\alpha^2+\alpha-1}{1+\alpha+\alpha^2}}
\end{equation}
One can easily verify that the maximal current (\ref{answer3},\ref{answer4}) exceeds the maximal possible rectification current at the same voltage $V$ in a linear wire \cite{FSV}
$\sim \frac{e^2V}{h}{\rm max} ([V\tau_c]^{3g},[V\tau_c])$.

\section{Conclusions}

We have found the rectification current in the absence (\ref{3-no-time})  and  in the presence (\ref{answer1},\ref{answer2}) of the time-reversal symmetry in Y-junctions.
In all cases the maximal rectification current is greater than in a linear wire with the same interaction strength and bias voltage. In the absence of the time-reversal symmetry
the rectification current can be comparable with the total ac current $\sim e^2V/h$ for sufficiently strong interaction strength, i.e., it achieves its maximal possible order of magnitude. This reflects the fact that both electron interaction and time-reversal symmetry breaking facilitate rectification. For most values of parameters the rectification
current is a nonmonotonous function of the bias voltage.

Our calculations are valid in the vicinity of various fixed points in the low voltage regime. In a general case, a junction is controlled by a stable fixed point
at low voltages. For repulsive interaction of spin-polarized particles there is only one stable fixed point:
three disconnected wires.\cite{5} We found
a stronger rectification effect near that fixed point than for two weakly connected wires. However, the current is low for weakly connected wires. The diode effect is much stronger in the vicinity
of unstable fixed points. Thus, it is important to understand how to tune the system close to those fixed points.
Some of them may be tricky to realize experimentally. Indeed, as discussed above, negative elements in the matching matrix $M$ imply Andreev reflection and could be obtained in a hybrid normal-superconductor structure or in cold atom systems \cite{demler}.
At the same time, it is straightforward to make an
``island junction'' \cite{10}, Fig.~1, with positive matching matrix elements
\begin{equation}
\label{flower}
M=\left(
\begin{array}{ccc}
0& 1& 0\\
0& 0& 1\\
1& 0& 0
\end{array}\right).
\end{equation}
For example, one can use three line junctions between three quantum Hall systems. The role of impurities is played by three constrictions in the junctions.
Alternatively, one can use a single quantum Hall island confined between edges AB, CD and EF (Fig.~1). The rectification current scales as
\begin{equation}
\label{isl-curr}
I_r\sim U^3 V^{\frac{12g}{g^2+3}-2}
\end{equation}
at $V\sim V^*=(U\tau_c)^{\frac{g^2+3}{(3-g)(1-g)}}/\tau_c$.
Time-reversal symmetry is broken in such setup. Thus, the rectification current can be made comparable to the total ac current even in the low voltage regime. This is the main result of the paper.

Most of the time we ignored phase factors $\exp(i\alpha_l)$ in tunneling operators. As the example of three weakly connected wires shows, for special values
of the phases
the rectification effect is suppressed. It has the same order of magnitude for other values of the phases.

We considered the simplest example of a junction: 3 spin-polarized wires.
It would be interesting
to generalize our results to a system with spin. This may result in a more complicated behavior as the minimal model with spin includes 6 channels: two for each wire.
Still, we expect a similar physics. In particular, a time-reversal-invariant system with spin can be obtained from two copies of the ``island junctions'' with opposite spins and
chiralities. In the absence of the interaction between the copies, the problem reduces to a spin-polarized island junction. An actual realization, based, e.g., on a topological insulator, must involve interaction between opposite spins. We expect that such interaction does not change the qualitative picture.

\begin{acknowledgments}
We acknowledge funding for this project from NSF under grants number DMR-0544116 and
PHY05-51164 and the U.S. Department of Energy, grant number DE-SCOOO1556.
DEF thanks KITP for hospitality.
\end{acknowledgments}

\end{document}